# Ambipolar diffusion of photo-excited carriers in bulk GaAs

Brian A. Ruzicka, Lalani K. Werake, Hassana Samassekou, and Hui Zhao[a)]

*Department of Physics and Astronomy, The University of Kansas, Lawrence, Kansas 66045, USA*

The ambipolar carrier diffusion in bulk GaAs is studied by using an ultrafast pump-probe technique with a high spatial resolution. Carriers with a point-like spatial profile are excited by a tightly focused pump laser pulse. The spatiotemporal dynamics of the carriers are monitored by a time-delayed and spatially scanned probe pulse. Ambipolar diffusion coefficients are deduced from linear fits to the expansion of the area of the profiles, and are found to decrease from about 170 cm$^2$s$^{-1}$ at 10 K to about 20 cm$^2$s$^{-1}$ at room temperature. Our results are consistent with those deduced from the previously measured mobilities.

Transport of charge carriers in semiconductors is an important aspect of carrier dynamics and plays an essential role in many electronic applications. For a carrier system under thermal equilibrium with the lattice, the transport process is well described by the drift-diffusion equation. The mobility and the diffusion coefficient are related by the Einstein relation, and are both well related to microscopic quantities such as the mean-free time and mean-free path of carriers. Transport of photo-excited carriers is related to, but different from, the transport of electrons or holes. Since the interband optical excitation generates electron-hole pairs that are bound by Coulomb attraction, the pairs move in the semiconductor as a whole. Because the photo-excited electron-hole pairs are neutral, they do not drift under an applied electric field; they only diffuse from high-density to low-density regions. Such an ambipolar diffusion process is the primary process of excitation transfer, and plays an important role in opto-electronic devices. For example in a photovoltaic device, the absorbed photon energy is transferred by diffusion of the electron-hole pairs before the charge separation.

Studies of the ambipolar diffusion of photo-excited carriers can provide fundamental knowledge for opto-electronic applications and complementary information for electrical studies of the unipolar electron or hole transport. Previously, ambipolar diffusion has been studied by several optical techniques including transient grating, pump-probe, and photoluminescence. In transient-grating experiments, a periodic spatial distribution of carrier density is generated by the interference of two laser pulses from different angles. The decay of this periodic distribution is then detected by the diffraction of a third pulse, and the diffusion coefficient can be deduced from this decay.[1–6] In pump-probe measurements, a laser pulse is used to excite carriers, and their distribution is then detected by measuring the transmission or reflection of a second laser pulse, which is time delayed and spatially scanned.[7–14] Photoluminescence spectroscopies, both in the time-of-flight configuration[15,16] and the spatially resolved geometry,[17–24] have also been applied to study ambipolar diffusion. Ambipolar diffusion coefficients in a large number of semiconductor structures have been measured, and examples include bulks of silicon,[2,13] CdSSe,[3] CdSe,[4] GaSe,[5] and CdS,[7] and quantum wells of ZnSe,[21–23] ZnCdSe,[17] and InGaAs,[20] and graphene.[14] Among them, the most extensively studied are GaAs-based systems. However, most experimental studies have been performed on GaAs quantum-well samples.[8–11,15,16,24–28] Since the scattering rates of carriers in quantum-well samples are different from bulk crystals due to the change of the density of states caused by the quantum confinement, the diffusion coefficients in quantum-well samples are different from the bulk values. Furthermore, in quantum-well samples the interface between the quantum well and the barrier causes additional scattering mechanisms. In contrast to these extensive studies on quantum-well samples,[8–11,15,16,24–28] studies on bulk GaAs are rare.[6]

Here we present a systematic study of the ambipolar diffusion coefficient in bulk GaAs with lattice temperatures in the range of 10 to 300 K by using an ultrafast pump-probe technique with a high spatial resolution. Carriers are excited with a point-like spatial profile by a tightly focused pump pulse. By monitoring the expansion of the profile, we deduce the ambipolar diffusion coefficient, which decreases monotonically from about 170 cm$^2$s$^{-1}$ at 10 K to about 20 cm$^2$s$^{-1}$ at room temperature. Our results are consistent with those deduced from the Einstein relation using the mobilities of electrons and holes measured by electrical techniques.

The sample studied is a 400-nm GaAs layer on a glass substrate, as we have described previously.[29] In our experiment, carriers are injected via the interband excitation of a 100-fs pump pulse with a central wavelength of 750 nm (Fig. 1a). The pump pulse is obtained by frequency doubling the signal output of an optical parametric oscillator (OPO), by using a BBO crystal (Fig. 1c). The OPO is pumped by a Ti:Sapphire laser with a repetition rate of 80 MHz. The pump pulse is focused to the sample through a microscope objective lens with a spot size of 1.6 $\mu$m full width at half maximum (FWHM). Once injected, the carriers diffuse from high density to low density regions, causing an expansion of the carrier density profile. Due to the Coulomb interaction, the excited electrons in the conduction band and positively charged holes in the valence band diffuse as pairs (Fig 1b). With a Gaussian initial profile, the solution

---

[a)]Electronic mail: huizhao@ku.edu

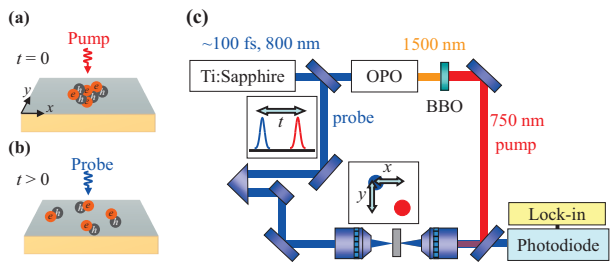

FIG. 1. Panel a: A tightly focused pump laser pulse injects pairs of electrons ($e$) and holes ($h$) in a bulk GaAs sample by interband absorption. Panel b: After a certain period of time, the injected carriers have diffused away from the excitation spot. The expanded carrier density profile is detected by a tightly focused probe pulse. Panel c: A schematic of the experimental set-up.

of the diffusion equation shows that the profile remains Gaussian, with a FWHM that increases with time as $w^2(t) = w_0^2 + 16\ln(2)D_a t$, where $D_a$ is the ambipolar diffusion coefficient.[8]

We study this ambipolar diffusion process by probing the carriers with a 100-fs probe pulse with a central wavelength of 800 nm. The probe pulse is taken from the output of the Ti:Sapphire laser before entering the OPO (Fig. 1c), and is focused to a spot of size 1.2 $\mu$m (FWHM) on the sample from the back side. The carrier density profile is measured by detecting the differential transmission $\Delta T/T_0 \equiv [T(N) - T_0]/T_0$, that is, the normalized difference between the transmissions with [$T(N)$] and without ($T_0$) carriers. It is straightforward to show, and we confirm experimentally, that $\Delta T/T_0 \propto N$, the carrier density.

In our measurements, we first acquire the carrier density profiles at different probe delays by scanning the probe spot in the x-y plane. Figure 2 shows the measured $\Delta T/T_0$ as a function of $x$ and $y$ for probe delays of 0, 10, 20, and 30 ps. Here $x = y = 0$ is defined where the centers of the pump and the probe spots overlap, and the probe delay $t = 0$ when the peaks of the pump and the probe pulses overlap. In this measurement, the pump pulse injects an average carrier density of about $10^{17}$ cm$^{-3}$, and the sample is cooled to 10 K. At $t = 0$, the Gaussian shape of the $\Delta T/T_0$ profile is consistent with the pump and probe laser spots, with a size determined by the convolution of the pump and probe spots. At later times, the profile remains Gaussian, as expected from the diffusion model, and becomes lower and wider due to the ambipolar diffusion.

Since no anisotropic diffusion is observed in Fig. 2, to quantitatively study the diffusion process and deduce the ambipolar diffusion coefficient, it is sufficient and more efficient to measure the cross sections of the profile on the $x$-axis for a large number of probe delays. Figure 3a shows the $\Delta T/T_0$ as a function of $x$ and $t$ with a $y = 0$. A few examples of the profile at different probe delays, along with the Gaussian fits, are plotted in Fig. 3b. By fitting all of the measured profiles, we deduce the squared width as a function of the probe delay, as shown as the symbols in Fig. 3c. A linear expansion of the squared width, as expected from the diffusion model, is clearly observed over the whole time range measured. From a linear fit, we deduce an ambipolar diffusion coefficient of about 170 cm$^2$s$^{-1}$.

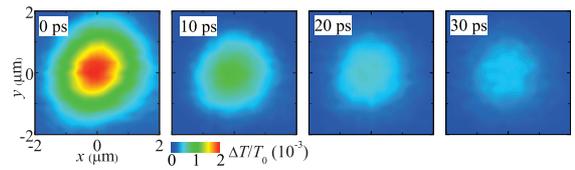

FIG. 2. Profiles of the differential transmission signal measured with probe delays of 0, 10, 20, and 30 ps, as labeled in each panel. The sample temperature is 10 K.

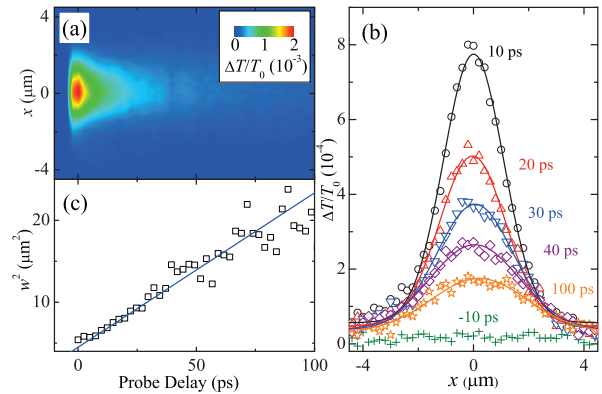

FIG. 3. Panel a: The differential transmission signal as functions of the probe delay and $x$, measured by scanning the probe spot along the $x$ axis and the probe delay. Panel b: cross section of Panel a at several probe delays as labeled in the figure. Panel c: the squared width of the profiles of the differential transmission signal as a function of probe delay obtained by fitting the profiles with Gaussian functions (solid lines in Panel b). The solid line is a linear fit, with a diffusion coefficient of 170 cm$^2$s$^{-1}$.

The procedure shown in Fig. 3 is then used to systematically study the ambipolar diffusion coefficient as a function of lattice temperature. The results are summarized in Fig. 4 (solid squares). At each temperature, multiple measurements at different sample locations were taken and the average value is plotted. Clearly, the ambipolar diffusion coefficient decreases monotonically with temperature over the range studied, from about 170 cm$^2$s$^{-1}$ at 10 K to about 20 cm$^2$s$^{-1}$ at room temperature.

It is worth mentioning that, due to the finite probe spot size, the profiles measured are actually convolutions of the probe spot and the actual carrier density profiles. However, since both the probe spot and the carrier density profiles are Gaussian, the convolution doesn't influence the measurement of $D_a$. Furthermore, this procedure of deducing $D_a$ is independent upon the carrier lifetime since the electron–hole recombination only influences the height, not the width, of the profiles.

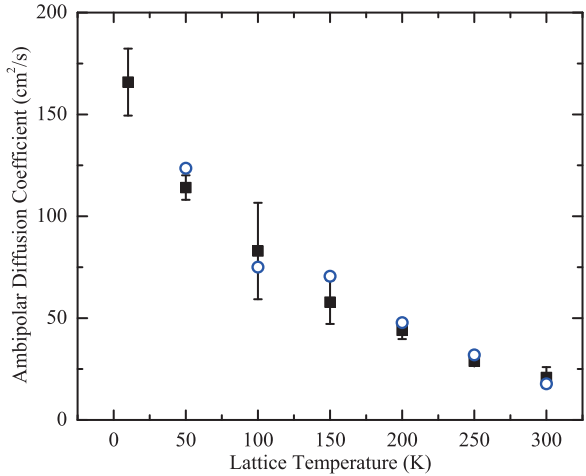

FIG. 4. The ambipolar diffusion coefficient as a function of the lattice temperature (solid squares) deduced by repeating the measurement shown in Fig. 3 with different temperatures. The open circles show the ambipolar diffusion coefficient calculated from reported mobilities of electrons and holes (see text).

Our results are reasonably consistent with a previous measurement for a temperature of 60 K (120 cm$^2$s$^{-1}$).[6] We are not aware of previous measurements of the ambipolar diffusion coefficient of bulk GaAs at other temperatures. However, it is interesting to compare our results with previously measured mobilities. The mobilities of electrons and holes in GaAs have been measured by many groups, and the results from high purity samples are reasonably consistent. Hole mobilities ($\mu_h$) measured in high-purity p-type GaAs at sample temperatures of 50, 100, 150, 200, 250, 300 K are 15, 4.5, 3, 1.5, 0.8, and 0.36 $\times 10^3$ cm$^2$V$^{-1}$s$^{-1}$.[30,31] Electron mobilities ($\mu_e$) have also been determined by using high-purity n-type samples, and the reported values are 32, 13, 3, 1.8, 1, 0.75 $\times 10^4$ cm$^2$V$^{-1}$s$^{-1}$ for these temperatures.[31,32] On the time scale of our measurements, carriers are in thermal equilibrium with the lattice. Hence, from these values, we can deduce the diffusion coefficients of electrons and holes ($D_e$ and $D_e$) for these temperatures by using the Einstein relation, $D_{e(h)}/k_B T = \mu_{e(h)}/e$, where $k_B$, $T$, and $e$ are the Boltzmann constant, the temperature, and the electron charge, respectively. We note that the reported values are Hall mobilities. Here we treat them as drift mobilities for simplicity. From the diffusion coefficients of electrons and holes, we deduce the ambipolar diffusion coefficients by using $D_a = 2D_e D_h/(D_e + D_h)$.[33] The results are plotted as the open circles in Fig. 4. Clearly, our results agree very well with these transport measurements.

In summary, we have studied ambipolar diffusion of photo-excited carriers in a bulk GaAs sample by using a high resolution pump-probe technique. Carriers with a point-like spatial profile are excited by a tightly focused pump laser pulse. The spatiotemporal dynamics of the carriers are monitored by a time-delayed and spatially scanned probe pulse. We found that the ambipolar diffusion coefficient decreases from about 170 cm$^2$s$^{-1}$ at 10 K to about 20 cm$^2$s$^{-1}$ at room temperature. Our results are reasonably consistent with those deduced from the previously measured mobilities by using the Einstein relation.

We thank John Prineas from University of Iowa for providing us with the GaAs sample. We acknowledge support from the US National Science Foundation under Awards No. DMR-0954486 and No. EPS-0903806, and matching support from the State of Kansas through Kansas Technology Enterprise Corporation.